\documentclass[showpacs,prl,floatfix,amsmath,amsfonts,superscriptaddress,preprint]{revtex4}
\usepackage{natbib,graphicx,amscd}
\usepackage[all,cmtip]{xy}
\usepackage[dvips]{changebar}
\usepackage[usenames]{color}
\usepackage[]{hyperref}
\newcommand{\pd}[2]{\frac{\partial #1}{\partial #2}}

\newcommand{\ud}{\mathrm{d}}

\begin{document}
\title{Model for a collimated spin wave beam generated by a
 single layer, spin torque nanocontact}
\date{\today}
\author{M. A. \surname{Hoefer}}
\email{hoefer@boulder.nist.gov}
\thanks{Contribution of the U.S. Government, not subject to copyright.}
\author{T. J. \surname{Silva}}
\affiliation{National Institute of Standards and Technology, Boulder,
  Colorado 80305, USA}
\author{M. D. \surname{Stiles}}
\affiliation{National Institute of Standards and Technology, Gaithersburg,
  Maryland 20899, USA}
\begin{abstract}
  % Magnetization dynamics are efficiently induced in nanostructures by
  % the exchange of angular momentum between carriers and a magnet.
  % While spin torque induced dynamics have been directly observed via
  % giant magnetoresistance in trilayer structures, %where two
  %   % ferromagnetic layers are separated by a nonmagnetic conductor, 
  % they have also been indirectly inferred for single layer structures.
  % However, little is understood about dynamics in the latter case.
  % Here, 
  A model of spin torque induced magnetization dynamics based upon
  semi-classical spin diffusion theory for a single layer nanocontact
  is presented.  The model incorporates effects due to the current
  induced Oersted field and predicts the generation of a variety of
  spatially dependent, coherent, precessional magnetic wave
  structures.  Directionally controllable collimated spin wave beams,
  vortex spiral waves, and localized standing waves are found to be
  excited by the interplay of the Oersted field and the orientation of
  an applied field.  These fields act as a spin wave ``corral'' around
  the nanocontact that controls the propagation of spin waves in
  certain directions.
\end{abstract}
\pacs{75.40.Gb %Dynamic properties (dynamic susceptibility, spin
               %waves, spin diffusion, dynamic scaling, etc.) 
  85.75.-d %Magnetoelectronics; spintronics: devices exploiting 
           % spin polarized transport or integrated magnetic fields 
  75.40.Mg %Numerical simulation studies 
  76.50.+g %Ferromagnetic, antiferromagnetic, and ferrimagnetic 
           %resonances; spin-wave resonance (see also 75.30.Ds Spin waves)
  75.30.Ds %Spin waves (for spin-wave resonance, see 76.50.+g)
  %74.78.Na %Mesoscopic and nanoscale systems 
  75.75.+a %Magnetic properties of nanostructures 
  75.70.Cn %Magnetic properties of interfaces (multilayers, superlattices, heterostructures) 
  72.25.Ba %Spin polarized transport in metals 
}

\maketitle

%=========================================================================

The flow of sufficiently large dc current through a thin, nanomagnetic
structure can give rise to precessional magnetization dynamics at
gigahertz frequencies \cite{Kiselev2003}.  This remarkable effect has
attracted broad interest, both from the standpoint of fundamental
physics and in the context of applications.  The underlying physics of
these spin torque devices is based upon the ability of thin
ferromagnetic layers to act as spin filters when current flows through
the layers.  For spin torque effects to manifest, a source of spin
polarized carriers with a component perpendicular to the magnetization
of a layer is required.  A typical spin torque multilayer has two
primary magnetic layers: a fixed layer that acts as a spin ``sieve''
that induces a spin accumulation in a non-magnetic spacer layer, and
an active layer that can respond dynamically when it absorbs the
angular momentum from the accumulated spins.  If the active layer has
a uniform magnetization, a torque is produced only when the two layers
are misaligned \cite{Slonczewski1996}.

On the other hand, if the magnetization is not uniform, theory for
even a single magnetic layer predicts a non-zero torque with resulting
dynamics \cite{Polianski2004,Stiles2004}.  Single layer spin torque
theory was used to explain differential resistance data in mechanical
nanocontact experiments \cite{Ji2003} and in lithographically defined
nanopillars \cite{Ozyilmaz2004}.  Theoretical studies have considered
single layer nanocontact devices \cite{Adam2006}, but have not
addressed the response of a physically realistic, finite sized
nanocontact with its accompanying Oersted field generated by the dc
current flowing through the device.

In general, the total spin accumulation, and hence spin torque, in a
magnetic thin film device arises from the lateral and longitudinal
diffusion of spins, transverse and parallel to the current direction,
respectively \cite{Stiles2004}.  The oft used Slonczewski model of
spin torque in trilayers, however, assumes a uniform spin accumulation
and incorporates only longitudinal spin diffusion effects
\cite{Slonczewski1996}.  

In this paper, we report a realistic treatment of a single layer
device including lateral and longitudinal spin diffusion, the Oersted
field, and a large enough sample geometry to capture novel behavior.
Using a novel micromagnetic simulator, we demonstrate unexpected
features of the response, including localized standing waves, vortex
spiral waves, and, most strikingly, a weakly diffracting collimated
beam of spin waves, the direction of which can be steered by changing
the direction of an applied magnetic field.  The formation of the beam
appears to be a novel physical mechanism involving the hybridization
of a localized standing wave and a vortex spiral wave.  It has been
previously shown that spin waves emitted from nanocontact devices can
be used to phase-lock two spin torque oscillators \cite{Kaka2005}.
The ability to steer a spin wave beam with magnetic field could offer
a method to selectively control phase locking of multiple spin torque
oscillators in an array structure.  

The outline of this paper is as follows.  First, we describe a
two-dimensional (2D) model of spin torque in single layer, nanocontact
devices.  We then present micromagnetic simulations that demonstrate
the wide variety of responses mentioned above.  Finally, we explain
the results of the simulations using a local formulation of the linear
spin wave dispersion relation above and below the nanocontact, showing
how the applied and Oersted fields act as a spin wave ``corral''
enabling directional control of anisotropic spin wave propagation and
localized excitations.  We also show that this corralling effect is
limited neither to this particular model nor the details of spin
torque.

\begin{figure}
  \centering
  \includegraphics[width=0.4\columnwidth]{%figures/
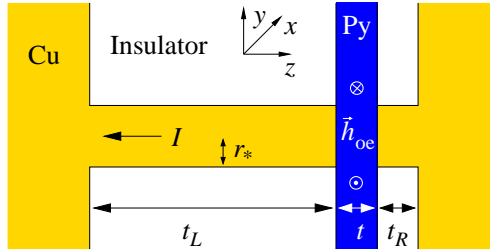}
  \caption{Single layer nanocontact device schematic.  The spatially
    nonuniform Oersted field $\vec{h}_{\textrm{oe}}$ is generated by
    the dc current $I$.  ``Py'' = Permalloy (Ni$_{80}$Fe$_{20}$).}
  \label{fig:schematic}
\end{figure}
The physical system we analyze is pictured in Fig.\
\ref{fig:schematic} and is similar to the one theoretically studied in
\cite{Stiles2004} except that we explicitly treat the finite contact
area and Oersted fields.  A single ferromagnetic Ni$_{80}$Fe$_{20}$
(Permalloy, Py) layer is adjacent to two copper (Cu) leads and an
insulator.  The current flows uniformly in the $-\hat{z}$ direction
(electron flow is in the $+\hat{z}$ direction) from a right reservoir
located at $z=t_R+t$, through the cylindrical Cu lead of radius $r_*$,
across the Py layer of thickness $t$ at $z=0$, over to a left
reservoir at a distance $t_L$ away from the magnetic layer ($z=-t_L$).
The length $t_R$ is an effective distance over which the current is
assumed to maintain quasi-unidirectional flow.  The magnetic layer is
assumed to have infinite extent in the $xy$ directions.  When we refer
to the region \emph{above} (\emph{below}) the nanocontact, we mean the
positive (negative) direction along the $y$-axis.

We calculate the spin accumulation due to current flow through a
ferromagnet using the same method described in Ref.\
\cite{Stiles2004}.  
% The spin accumulation due to current flow through
% a ferromagnet in the limit of small amplitude magnetic excitations.
We consider the behavior of the spin accumulation in the non-magnet in
response to a non-uniform magnetization $\vec{M}= M_{\rm s} \vec{u}$,
where $M_{\rm s}$ is the saturation magnetization.  As in Ref.\
\cite{Stiles2004}, we treat the case of small deviations
$\vec{u}_\perp$ away from the spatially averaged ``equilibrium''
direction $\vec{u}_\parallel$.  The transverse and longitudinal
components of the spin accumulation $\vec{m} = \vec{m}_\perp +
\vec{m}_{\parallel}$, where $\vec{m}_{\parallel}$ points in the
longitudinal direction of the steady state spin accumulation in the
absence of any magnetic inhomogeneity, can be decoupled in this limit.
We solve the multipoint boundary value longitudinal problem to find
the longitudinal spin accumulation $m_\parallel$ and spin current
$Q_{zz}$ for each interface.  For deviations at a particular
transverse wave vector $(k_x,k_y)$, the transverse spin accumulation
is then given in terms of the deviations in the magnetization and the
longitudinal solution
\begin{equation}
  \label{eq:1}
  \begin{split}
    \mathcal{F}\{\vec{m}_{\perp}\} = &\mathcal{F}\{\vec{u}_{\perp}\}
    \frac{\pm Q_{zz} + w_0 m_\parallel}{D \kappa \coth(l' \kappa) + w_0} , \\
    &\kappa = (k_x^2 + k_y^2 + 1/l_{sf}^2)^{1/2},
  \end{split}
\end{equation}
where $\mathcal{F}\{\vec{m}_{\perp}\}$ and
$\mathcal{F}\{\vec{u}_{\perp}\}$ are the 2D Fourier transforms of the
transverse spin accumulation and the magnetization transverse to the
average, respectively ($+$ for right $z=t$ interface and $-$ for left
$z=0$ interface).  The decoupling of the longitudinal spin
accumulation from the transverse spin accumulation is strictly valid
only in the limit of small deviations from a uniform magnetization
distribution.  In our case, the deviations from uniformity are not
small, so this treatment should be considered to be a first order
approximation.
% A more rigorous treatment
% is beyond the scope of this work.  
The distance to the reservoir $l'$ is $t_L$ or $t_R$, for the left and
right interfaces respectively.  The spin diffusion length $l_{sf}$ and
diffusion constant $D$ are material parameters for Cu; $w_0$ is the
effective interfacial spin absorption rate.

To calculate the transverse spin accumulation in real, polar
coordinate $(r,\phi)$ space, we take the inverse Fourier transform of
eq.\ \eqref{eq:1}.  Then the expression for $\vec{m}_{\perp}$, eq.\
\eqref{eq:16} (see appendix), is the convolution of the magnetization
with a weakly singular kernel over the point contact region.  This
nonlocal formulation of the transverse spin accumulation can be
interpreted as the lateral diffusion of spins interacting with a given
magnetization distribution.

We find the quasi-steady-state spin accumulation for a given
instantaneous magnetization distribution in the Py layer.  This is
justified because the ratio of the time scales for the diffusion of
electrons to a steady state and for the magnetization dynamics is
about 0.001.
% so that transient spin dynamics in the Cu
%leads have a negligible effect on the magnetization.  
By formulating the calculation of the inhomogeneous transverse spin
accumulation in terms of a simple convolution operation, we have
greatly improved the speed of simulating this effect as compared to
directly calculating the coupled magnetization and spin accumulation
\cite{Garcia-Cervera2007}.

We use the average magnetization direction over the contact
$\hat{u}_\parallel$, eq.\ \eqref{eq:13}, as the orientation of the
longitudinal spin accumulation.
% in order to calculate the \emph{total} spin accumulation.
Physically, this corresponds to the situation where a spin scattered a
large number of times from the interface effectively ``sees'' the
average magnetization.  The total spin accumulation inside the
nanocontact $\vec{m}^*$, eq.\ \eqref{eq:14}, is then calculated by
summing the contributions to the longitudinal and transverse
accumulation from each interface.  Because an insulator surrounds the
Cu leads in our model, the spin accumulation outside the nanocontact
is zero.
%$\vec{m}^*(r,\phi,\tau) = 0$, $r>1$.

The dynamical equation for the magnetization is
\begin{equation}
  \label{eq:6}
  \begin{split}
    \frac{\partial \vec{u}}{\partial \tau} &= - \vec{u} \times
    \vec{h}_{\text{eff}} - \alpha \vec{u} \times (\vec{u} \times
    \vec{h}_{\text{eff}}) + \sigma \vec{u} \times (\vec{u} \times
    \vec{m}^* ), \\
    \vec{h}_{\text{eff}} &= \vec{h}_0 - u_z \hat{z} - g(r) \hat{\phi}
    + \eta \nabla^2 \vec{u} , ~ \sigma = \frac{\hbar w_0}{2 t \mu_0
      M_{\rm s}^2}.
  \end{split}
\end{equation}
This is a modified 2D Landau-Lifshitz equation in dimensionless form
with time normalized by $\gamma \mu_0 M_{\rm s}$ ($\gamma$ is the
gyromagnetic ratio, $\mu_0$ the permeability of free space).  In eq.\
\eqref{eq:6}, space is normalized by $r_*$, fields and magnetization
are normalized by $M_{\rm s}$, $\alpha$ is the damping constant, $\eta
= D_{\rm ex}/(\gamma \mu_0 M_{\rm s} \hbar r_*^2)$ is the coefficient
of the exchange term ($D_{\rm ex}$ is the exchange parameter, $\hbar$
is Planck's constant divided by $2\pi$), $\vec{h}_0 =
h_0[\sin(\theta_0)\cos(\psi_0),\sin(\theta_0)\sin(\psi_0),\cos(\theta_0)]$
represents the canted, normalized applied field ($\theta_0$ is
measured from the positive $z$-axis, $\psi_0$ is measured
counterclockwise from the positive $x$-axis), $-u_z \hat{z}$ is the
axial demagnetizing term, and $\vec{h}_J(r,\phi) = -g(r) \hat{\phi}$
is the nonuniform Oersted field due to the current density $J(r)$
defined below.
% , calculated in \cite{Hoefer2006a}.  
The driving torque is similar to the Slonczewski torque for a trilayer
device \cite{Slonczewski1996} except that here, the torque is
nonlocal.  We have only considered the leading order axial component
of the dipole field, $-u_z \hat{z}$ that is independent of film
thickness $t$.  Calculation of higher order terms involves integration
over the ferromagnetic volume that greatly complicates the solution of
eq.\ \eqref{eq:6}.  Local approximations to higher order correction
terms for such dipole fields have been derived by Arias and Mills
\cite{Arias2007}.  Such correction terms scale as $t \nabla \cdot
\vec{u}$ and have quantitative significance for large film
thicknesses.  
% The approximation made here is strictly valid in the
% limit of large exchange energy relative to the dipole energy
% associated with these correction terms, or $t r_* \ll 2D_{\rm
%   ex}/\hbar \gamma \mu_0 M_{\rm s}$.  Because $t r_* \approx 600
% ~\textrm{nm}^2$ and $2D_{\rm ex}/\hbar \gamma \mu_0 M_{\rm s} \approx
% 66 ~ \textrm{nm}^2$ for the simulation parameters considered here, it
% is expected that inclusion of such dipolar correction terms will
% affect the quantitative results.

The Oersted field is significant in the nanocontact geometry, with a
maximum magnitude on the order of 80 kA/m (1000 Oe) for $r_* = 40$ nm
and $I = 21$ mA.  Thus, to ignore the Oersted field, as was done in
previously presented multilayer simulations \cite{Hoefer2005}, is not
an appropriate approximation.  We will demonstrate that the Oersted
field significantly affects the response of the system.  Now we derive
the Oersted field for the geometry shown in Fig.\ \ref{fig:schematic}.

The current density $J(r',z')$ in dimensional units $(r',z')$ is
modeled as
\begin{equation}
  \label{eq:2}
  \begin{split}
    &\vec{J}(r',z') = \\
    & \hat{r} \frac{I}{2\pi D_p r_*} F(r') 
     [H(z'-t-t_R)H(t+t_R+D_p-z') - \\
    & \qquad \qquad \quad \,\; H(-t_L-z')H(z'+t_L+D_p)] \\ 
    &  - \hat{z} \frac{I}{\pi r_*^2}
    H(r')H(r_*-r')H(t+t_R-z')H(z'+t_L) , \\
    &F(r') = \left\{
      \begin{array}{cc}
        r'/r_* & 0 < r' < r_* \\
        r_*/r' & r_* < r'
      \end{array} 
    \right . \!\!\!, \quad 
    H(x) = \left\{
    \begin{array}{cc}
      0 & x < 0 \\
      1 & x > 0
    \end{array} \right. .
  \end{split}
\end{equation}
% This current density models the situation depicted in
% Fig. \ref{fig:schematic}.
Two infinite parallel conductor plates of thickness $D_p$, separated a
distance $t_R+t+t_L$,
% are assumed to have infinite extent because
% the experimental leads are more than an order of magnitude larger than
% the point contacts.  These plates
are connected by a wire of radius $r_*$.  Current flows into the wire
from the right plate, through the magnetic layer via a cylinder
modeling the point contact, and out of the wire into the left plate.
The current flow in the plates is assumed to be in the radial
direction
%$\hat{r}$ 
only.  
%The distance between the plates is $A$.  
The coefficient of $\hat{r}$ in \eqref{eq:2} models the magnitude of
the current density in the plates, assumed uniform in the $z$
direction.  Outside of the region where the wire connects to the
plates ($r > r_*$), the current density falls off proportional to
$1/r$, whereas inside the region ($r < r_*$), the current is
proportional to $r$.  The current is assumed to flow uniformly down
the wire modeled by the coefficient of $\hat{z}$ in \eqref{eq:2}.
This model of current flow conserves the flux of current from the
plates to the wire.  We are interested in the value of the Oersted
field $\vec{H}_J$, due to the current density (\ref{eq:2}), in the
center of the magnetic layer $z = t/2$.

Using a vector potential representation along with Fourier and Hankel
transforms, we solve for the Oersted field in dimensionless units
$(r,z) = (r'/r_*,z'/r_*)$ \cite{Hoefer2006a}:
\begin{align}
  \vec{h}_J(r,\phi) = -\hat{\phi} \frac{I}{M_{\rm s} \pi r_*} \{ &
  \underbrace{I_{1}(\frac{r}{r_*})}_{\text{infinite wire}} -
  \underbrace{I_{2}(\frac{r}{r_*})}_{\text{finite wire correction}} +
  \nonumber \\
  \label{eq:3}
    &\underbrace{ \frac{r_*}{M_{\rm s} D_p} [ I_{3}(\frac{r}{r_*}) +
      I_{4}(\frac{r}{r_*})]}_{\text{conductive plates}} \} \\
    = -g(r) \hat{\phi} \,\, & \nonumber
\end{align}
The integrals in (\ref{eq:3}) are
\begin{equation}
  \label{eq:4}
  \begin{split}
    &I_{1}(\rho) \! = \! \left\{
      \begin{array}{cc}
        \rho/2 & 0 < \rho < 1 \\
        1/2\rho & 1 < \rho 
      \end{array}
    \right . , \\
    &I_{2}(\rho) \! = \! \int_0^\infty \! e^{-qa/2} \cosh(q
    z_*) J_1(q) J_1(q \rho) \frac{1}{q} \, \ud q, \\
    &I_{3}(\rho) \! = \! \int_0^\infty \! e^{-qa/2}(1 \! - \! e^{-qd})
    \cosh(q z_*) J_0(q) J_1(q \rho) \frac{1}{q} \, \ud q, \\
    &I_{4}(\rho) \! = \! \int_0^\infty \! e^{-qa/2}(1 \! - \! e^{-qd})
    \cosh(q z_*) J_2(q) J_1(q \rho) \frac{1}{q} \, \ud q ,
  \end{split}
\end{equation}
where $a = (t_R+t+t_L)/r_*$, $d = D_p/r_*$, $z_* = (t_R-t_L)/2r_*$ are
the normalized wire length, conductive plate thickness, and the
location of the magnetic layer, respectively.  $J_n$ are the
$n$th-order Bessel functions of the first kind.  Note that
(\ref{eq:3}), with $I_{2} = I_{3} = I_{4} \equiv 0$, is the result for
the magnetic field due to an infinitely long wire of radius $r_*$ with
no conductive plates.  The magnitude of the Oersted field $h_J(r) =
|\vec{h}_J(r,\phi)|$ in (\ref{eq:3}) depends only on $r$.

\begin{figure}[h]
  \centering
  \includegraphics[width=0.5\columnwidth]{%figures/
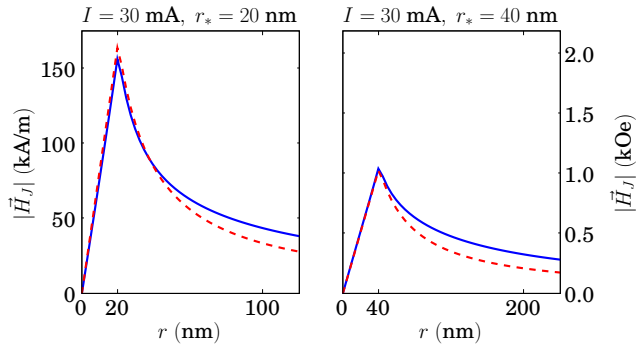}
  \caption{The magnitude of the Oersted magnetic
    field (\ref{eq:3}) (solid) and the Oersted field due to an
    infinitely long wire with no conductive plates (dashed).  
  }
  \label{fig:oersted_field}
\end{figure}

By numerically evaluating the integrals (\ref{eq:4}), we show $M_s
h_J(r)$ for two contact sizes and a fixed current $I = 30$ mA in Fig.\
\ref{fig:oersted_field}.  The fields due to an infinitely long wire
with no conductive plates are also shown in Fig.\
\ref{fig:oersted_field} as dashed curves.  Equation \eqref{eq:3} and
the result for an infinite wire agree near the point contact but the
decay of the fields is faster for the infinite wire case.

\begin{figure}
  \centering
  \includegraphics[width=0.5\columnwidth]{%figures/
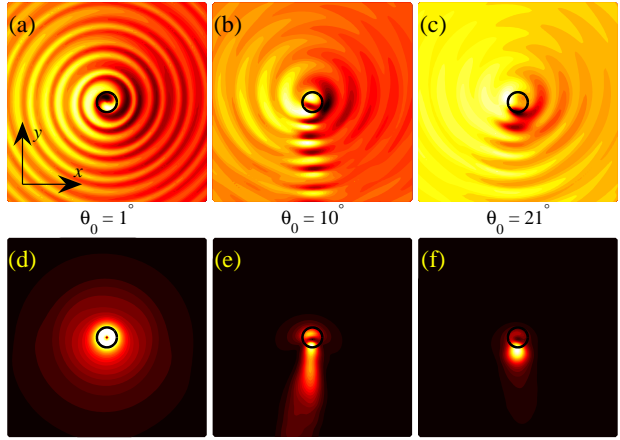}
  \caption{Magnetization ($u_y$) pattern and energy density of a thin
    film excited by dc current through a nanocontact with varied
    applied field canting angle $\theta_0 = 1^\circ, ~ 10^\circ, ~
    21^\circ$ in (a,d), (b,e), and (c,f), respectively.  All other
    system parameters are fixed ($I=29$ mA, $\psi_0 = 0$).  The circle
    in the center represents the boundary of the nanocontact.  The
    domain is a square ten times the contact diameter per side.
    Magnitudes in each panel are normalized; positive values are
    yellow, negative values are black, with $u_y$ oscillating between
    approximately $\pm 0.8$, $\pm 0.6$, $\pm 0.65$ in (a), (b), and
    (c), respectively.  The peak energy density at $r=10$ in (d), (e), and (f)
    is, in arbitrary units, $0.74$, $1.0$, and $0.15$
    % $0.017$, $0.023$, and $0.0035$,
    respectively.  Note the pinning of the vortex core in (a) and (d).
    The energy density plot in (e) clearly shows the weak diffraction
    of the spin wave beam.}
  \label{fig:modes_angle}
\end{figure}
We implemented a numerical method to solve eq.\ \eqref{eq:6} in polar
coordinates, (see appendix).
% details of which will appear in a future publication.
The calculations are rendered tractable by formulating the model in a
nonuniform polar coordinate grid, allowing us to compute over a large
domain (4.8 $\mu$m diameter disk) to avoid boundary spin wave
reflections and with simulation times (3 ns) sufficiently long to
ensure that we have determined the true steady state response.  By
evolving eq.\ \eqref{eq:6} in time with a nonuniform initial condition
(where $\vec{u}$ is relaxed in the presence of the effective field
only), we find that the magnetization settles into a quasiperiodic
state due to the competition between the spin accumulation torque and
the damping.  All excitation frequencies are calculated from the time
series of $u_y$ averaged over the nanocontact by use of Fourier
methods with a typical resolution of 0.75 GHz.  We use the physical
parameters listed in Table I of \cite{Stiles2004} except: $h_0 = 1.1$,
$r_* = 40$ nm, $t_R = 5$ nm, $t_L = 75$ nm, $t = 15$ nm, $M_{\rm s} =
800$ kA/m, $D' = 0.001 ~ \textrm{m}^2/\textrm{s}$ (diffusion rate in
Py), $l_{sf}^{FM} = 5.5$ nm (spin diffusion length in Py), and $D_p =
50$ nm.  
% (conductive plate thickness in Oersted field model
% \cite{Hoefer2006a}).

\begin{figure}
  \centering
  \includegraphics[width=0.5\columnwidth]{%figures/
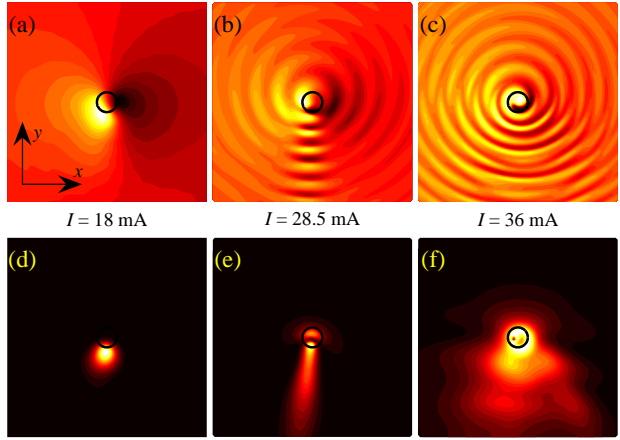}
  \caption{Magnetization ($u_y$) pattern and energy density of a thin
    film excited by dc current through a nanocontact with varied dc
    current $I = 18, ~ 28.5, ~ 36$ mA in (a,d), (b,e), and (c,f),
    respectively.  All other system parameters are fixed ($\theta_0 =
    10^\circ$, $\psi_0 = 0^\circ$).  $u_y$ oscillates between
    approximately $[-0.37,0.34]$, $[-0.53,0.52]$, $[-0.97,0.87]$ in
    (a), (b), and (c), respectively.  The peak energy density at
    $r=10$ in (d), (e), and (f) is, in arbitrary units, $0.0051$,
    $0.35$, and $1.0$
    % $0.00028$, $0.019$, $0.055$
    respectively.}
  \label{fig:modes_I}
\end{figure}
Our calculations show a variety of behaviors that depend on the
physical parameters including vortex spiral waves (Figs.\
\ref{fig:modes_angle}a, \ref{fig:modes_I}c), spin wave beams (Figs.\
\ref{fig:modes_angle}b, \ref{fig:modes_I}b), and localized standing
waves (Figs.\ \ref{fig:modes_angle}c, \ref{fig:modes_I}a).  The
direction of spin wave propagation in a canted field, $\theta_0 > 0$,
depends on the applied field azimuthal angle $\psi_0$ where the spin
wave angle $\approx \psi_0 - 90^\circ$. The top panels depict the
spatial variation of $u_y$ at a specific time and the lower panels
show the energy density for each state.  The energy density
$E(r,\phi)$ is calculated by taking the time average of the squared
magnitude of the magnetization transverse to the average direction
$E(r,\phi) = \left< |\vec{u} \times \left<\vec{u} \right>|^2 \right>$,
where $\left< f \right> = \frac{1}{T} \int_{\tau_*}^{\tau_*+T} f \ud
\tau$, and $\tau_*$ is the time by which the magnetization has settled
into a steady, precessional state.

Our choice of a relatively thick Py layer emphasizes the effect of the
Oersted field.  Larger currents are necessary to excite thicker
layers, so that the Oersted fields are larger.  Calculations with a
thinner Py film still result in localized standing waves, spiral
waves, and anisotropic waves, but the collimated beam is more
difficult to excite.

The particular response excited can be explained, in part, by
appealing to the dispersion relation for a spin wave propagating in a
canted uniform field, eq.\ \eqref{eq:5}.  The local fields in the
regions above and below the nanocontact (on either side of the
nanocontact in the plane of the magnetic film) are of different
magnitudes and orientations due to the presence of the nonuniform
Oersted field $-g(r) \hat{\phi}$.  The Oersted field acts as a
``corral'' and effectively lifts the spatial degeneracy of the
dispersion relation immediately above and below the nanocontact so
that, at a given frequency, the spin waves propagate in one direction
and evanesce in the other.

\begin{figure}
  \centering
  % Figure size in Matlab is 8.65 cm x 3 cm
  \includegraphics[width=0.5\columnwidth]{%figures/
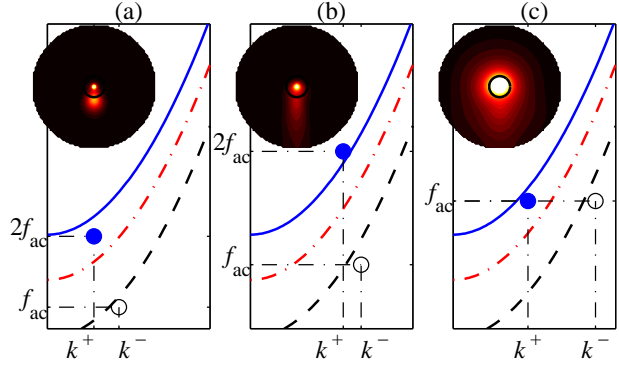}
  \caption{Band structure, excited frequencies and response energy
    density for three different driving frequencies (a) $f_{\rm ac} =
    10$ GHz, (b) $f_{\rm ac} = 16$ GHz, (c) $f_{\rm ac} = 25$ GHz.
    The filled/hollow circles correspond to the numerically determined
    wavenumber above/below the nanocontact and agree with the linear
    dispersion relation.  The FMR frequencies above and below the
    nanocontact are $f_{\rm FMR}^+ = 20.2$ GHz and $f_{\rm FMR}^- =
    5.6$ GHz, respectively.  When the drive frequency is in the gap
    between the upper and lower dispersion relations, the second
    harmonic $2f_{\rm ac}$ is also excited.}
  \label{fig:bands}
\end{figure}

The response away from the nanocontact does not strongly depend on the
details of spin torque except that the spin torque is localized at the
nanocontact.  To show this, we solved eq.\ \eqref{eq:6} with a
localized ac applied field \{$\vec{h}_{\rm ac}(r,\phi,t) = h_{\rm
  ac}\sin(2\pi f_{\rm ac} \tau) \hat{z}$, $r \le 0.15$, $0$
elsewhere\}, neglecting spin torque.  The response and the associated
dispersion curves are depicted in Figs.\ \ref{fig:bands}a-c for $I =
30$ mA, $\theta_0 = 10^\circ$, $\psi_0 = 0^\circ$, and $h_{\rm ac} =
1$.  We use the dispersion relation with the local fields evaluated at
$r = 1$ to approximate which wavenumbers can propagate above (solid
blue) and below (dashed black) the nanocontact.  The dash-dotted red
curve is the far field dispersion curve where the Oersted field is
negligible.  The Oersted field creates a gap between the dispersion
curves above and below the nanocontact.  The type of response excited
depends on the driving frequency $f_{\rm ac}$ and its relation to the
ferromagnetic resonance (FMR) frequencies (eq.\ \eqref{eq:5} with no
exchange contribution, $\eta = 0$) above ($f_{\rm FMR}^+$), below
($f_{\rm FMR}^-$), and far away ($f_{\rm FMR}^{\rm far}$) from the
nanocontact.  When the driving frequency is between $f_{\rm FMR}^-$
and $f_{\rm FMR}^{\rm far}$, the excitation is a standing wave (see
Fig.\ \ref{fig:bands}a).  When the driving frequency is between
$f_{\rm FMR}^{\rm far}$ and $f_{\rm FMR}^+$, the excitation forms a
spin wave beam (see Fig.\ \ref{fig:bands}b).  When the driving
frequency is above $f_{\rm FMR}^+$, the excitation is a nonlocalized
propagating wave (see Fig.\ \ref{fig:bands}c).  This is the ``corral''
effect; the applied field and the Oersted field act in concert to
modify the availability of spin wave states in close proximity to the
nanocontact, depending upon the excitation frequency.  In the
locations where the Oersted and applied fields add (above the
nanocontact), the dispersion relation is shifted upward in frequency,
thereby acting as a ``fence'' to block spin wave propagation if the
excitation frequency is below $f_{\rm FMR}^+$. Where the Oersted and
applied fields cancel, the dispersion relation is shifted downward in
frequency, thereby acting as a trap, localizing the spin wave
excitations until the excitation frequency exceeds $f_{\rm FMR}^{\rm
  far}$, whereupon the trap is no longer operative, and the spin waves
are guided downward.  The mechanism leading to the standing wave is a
linear phenomenon, in contradistinction to previous calculations where
nonlinear effects lead to the formation of a localized magnetic
excitation in a nanocontact geometry \cite{Slavin2005a}.  Even in the
presence of a small driving field (e.g., $h_{\rm ac} = 0.001$), both
the localized wave and the spin wave beam can be excited, precluding
any strong role of nonlinearity in the localization of the response.

\begin{figure}
  \centering
  \includegraphics[width=0.5\columnwidth]{%figures/
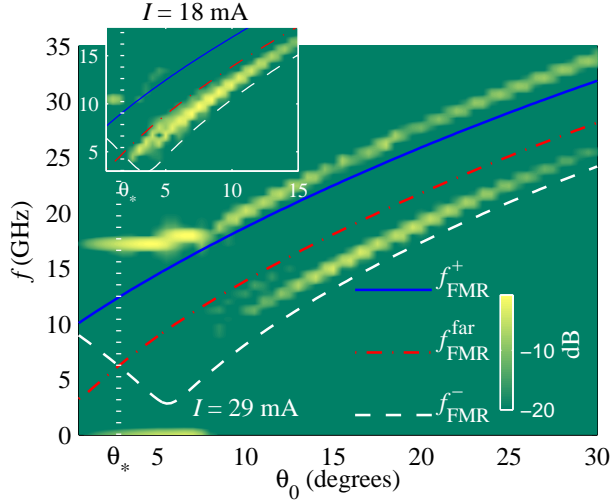}
  \caption{Nanocontact frequency spectrum as a function of applied
    field polar canting angle $\theta_0$ along with the local FMR frequencies
    above ($f_{\textrm{FMR}}^+$), below ($f_{\textrm{FMR}}^-$), and
    far away ($f_{\textrm{FMR}}^{\textrm{far}}$) from the nanocontact.
    % are calculated by evaluating the Oersted and applied fields above,
    % $(x,y) = (0,1.5)$, below, $(x,y) = (0,-1.5)$, and far from, $x^2 +
    % y^2 \to \infty$, the nanocontact.  
    Main plot current: $I = 29$ mA; inset plot current: $I = 18$ mA.}
  \label{fig:fmr_angle}
\end{figure}

\begin{figure}
  \centering
  \includegraphics[width=0.5\columnwidth]{%figures/
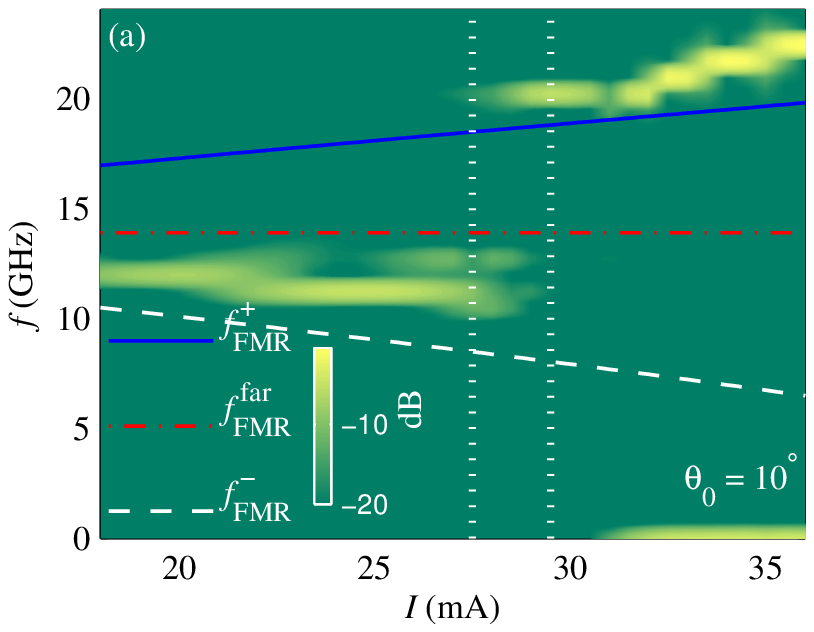}
  \\
  \includegraphics[width=0.5\columnwidth]{%figures/
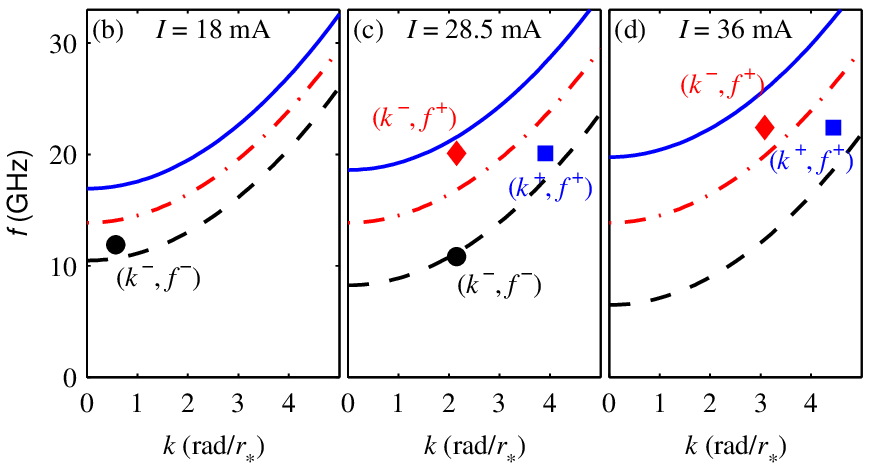}
  \caption{(a) Nanocontact frequency spectrum as a function of dc
    applied current. (b-d) Local dispersion curves evaluated above,
    (solid blue curve), below (dashed black curve), and far from
    (dash-dotted red curve) the nanocontact along with numerically
    calculated wavenumbers and frequencies.  (b) The circle is the
    wavenumber, frequency pair $(k^-,f^-)$ calculated from the
    localized wave in Fig.\ \ref{fig:modes_I}a.  (c) The diamond
    (square) corresponds to the wavenumber calculated above (below)
    the nanocontact $k^-$ ($k^+$) in Fig.\ \ref{fig:modes_I}b.  The
    two frequencies $f^-$ and $f^+$ are peaks in the spectrum at $I =
    28.5$.  (d) The diamond and square correspond to the wavenumber,
    frequency pairs as discussed for (c) but at the current $I = 36$
    mA in Fig.\ \ref{fig:modes_I}c.}
  \label{fig:fmr_current}
\end{figure}

We can interpret the different spatial responses in Figs.\
\ref{fig:modes_angle}, \ref{fig:modes_I} by plotting the excitation
spectrum as a function of applied field angle (Fig.\
\ref{fig:fmr_angle}) and applied dc current (Fig.\
\ref{fig:fmr_current}).  All future local FMR and dispersion curve
calculations assume the Oersted field $-g(r) \hat{\phi}$ is evaluated
at the radius $r = r_0 = 1.5$.  For currents close to the excitation
threshold, the response spectrum has a single frequency peak that
corresponds to vortex spiral waves for small angles and a localized
standing wave for larger angles (see the inset of Fig.\
\ref{fig:fmr_angle}).  The lowest available frequency response is
excited.  For small angles, the localized wave is not available
because $f_{\rm FMR}^{\rm far}$ (red, dash-dotted curve) is less than
$f_{\rm FMR}^{-}$ (white, dashed curve) so spin waves can propagate to
the far field and the vortex spiral wave is excited.  Above the
critical angle $\theta_* = \sin^{-1}[g(r_0)/(2h_0)]$ where $f_{\rm
  FMR}^{\rm far} = f_{\rm FMR}^{-}$ at the radius $r=r_0$, the
localized wave with a much lower frequency is excited because its
frequency is below $f_{\rm FMR}^{\rm far}$ (see the vertical dotted
lines in Fig.\ \ref{fig:fmr_angle}).

For larger applied currents, nonlinear effects appear to be important.
In Fig.\ \ref{fig:fmr_angle}, the vortex spiral wave is excited for
small angles for the same reasons mentioned previously.  However, the
bifurcation in mode behavior occurs for angles larger than $\theta_*$
where, now, there are two distinct frequency peak branches in the
spectrum that are not harmonically related.  These branches lie just
above $f_{\rm FMR}^{-}$ and $f_{\rm FMR}^{+}$ (solid, blue curve).  In
the small amplitude case (Fig.\ \ref{fig:fmr_angle} inset), these
branches correspond to the localized standing wave and the vortex
spiral wave, respectively.  Here, nonlinearity appears to spawn a
hybridization of the standing wave and the vortex spiral wave
resulting in the spin wave beam (see Figs.\ \ref{fig:modes_angle}b and
\ref{fig:modes_I}b).  As the angle is increased, the power associated
with each branch changes from being predominantly in the vortex spiral
wave branch to being predominantly in the localized wave branch and
back again resulting in a visual change of the energy density as shown
in Fig.\ \ref{fig:modes_angle}c where the standing wave is dominating
the response.

This nonlinear hybridization is further investigated in Fig.\
\ref{fig:fmr_current} where the frequency spectrum is plotted as a
function of applied current (a) and associated local spin wave
dispersion curves are plotted (b-d).  For small currents, the
localized standing wave is excited due to the corral effect (see Fig.\
\ref{fig:fmr_current}b and compare with Fig.\ \ref{fig:bands}a).  For
large currents, anisotropic waves are excited because the corral is
partially open (see Fig.\ \ref{fig:fmr_current}d ).  For intermediate
currents, between the vertical dashed lines, both the localized wave
and the spiral wave are excited at the same time, but with different
frequencies $f^-$ and $f^+$.  It is only within this current range
that we observe the spin wave beam.  There is a \emph{single}
wavenumber associated with these two frequencies, $k^-$, if the
dispersion curve below (above) the contact is associated with the
localized standing wave (vortex spiral wave) (see Fig.\
\ref{fig:fmr_current}c).  The spin wave beam corresponds to a
wavenumber $k^+$ associated with the dispersion curve below the
contact but with the larger, vortex spiral wave frequency $f^+$.
These two frequencies and two wavenumbers form a triad in the
dispersion diagram \ref{fig:fmr_current}c.  We interpret this behavior
in the following way:
\begin{equation*}
  \left. \begin{array}{c} 
      \xymatrixrowsep{1pc}
      \xymatrix{
      (f^+,k^-) ~ \text{spiral} \\
       \\ 
      (f^-,k^-) ~ \text{localized} \ar@{=>}[uu]|-{\textstyle \text{nonlinearity}}
    }
  \end{array} 
  \right \}
  \xymatrixcolsep{5pc}
  \xymatrix{
    \ar@{=>}[r]_-{\textstyle \text{hybridization}} & (f^+,k^+) ~
    \text{beam} .
  }
\end{equation*}
Given the coarse approximations made in this analysis (linear, 1D
behavior and pointwise evaluation of the Oersted field, etc.), the
discussion provides a qualitative description of the response
selection in single layer nanocontacts.

The results predicted here for nanocontacts stand in stark contrast
to calculations involving nanopillar structures \cite{Krivotorov2007}
where excitations are confined to a nanomagnet.  In the latter case,
the Oersted field was shown to effect a small perturbation to the
magnetization dynamics.  Here, the Oersted field gives rise to the
``corral'' effect and is fundamental to the formation of magnetic
excitations.

We have also performed numerical simulations in trilayer structures
(ferromagnet/nonmagnet/ferromagnet) neglecting lateral diffusion and
nonlocal effects and using the Slonczewski torque
\cite{Slonczewski1996} in both our own simulator and in a conventional
micromagnetic simulation package (OOMMF) by a colleague \cite{Russek}.
We find similar qualitative behavior such as the corralling effect and
a spin wave beam.  However, we do not observe stable vortex spiral
waves, and the spatial dependence of the excitations are different.
For spin torque magnitudes comparable to that of the single layer
torque discussed in this work, an unphysical, large spin transfer
efficiency of 0.7 was required.  Theoretical calculations suggest that
the spin transfer efficiency is between 0.25 and 0.4
\cite{Slonczewski1996,Pufall2003}.  Thus, we find that the torque
generated by a single layer with nonuniform magnetization can exceed
that expected for a uniformly magnetized layer in a trilayer
structure, suggesting that inclusion of spin diffusion effects
considered here in trilayer calculations may have significant
quantitative impact.  We note that the semiclassical transport
calculations in \cite{Stiles2004} used here do not require the use of
the spin transfer efficiency, an ad hoc parameter, because the spin
accumulation is directly calculated in an approximate but
self-consistent manner.  We also note that spin diffusion in a
trilayer structure has been studied in a self-consistent manner in
ref.\ \cite{Brataas2006}, but only the linear instability was
calculated (i.e.\ the threshold current for spin wave excitations),
and the calculations were strictly for the case of a uniform current
density flowing through an infinitely extended magnetic film without
any consideration of boundary conditions.  We are currently extending
our work to the case of a trilayer, nanocontact structure so that the
spin accumulation will be calculated in a self-consistent fashion,
precluding the need to invoke a spin efficiency factor.

In summary, we have used micromagnetics to predict the generation of a
variety of responses in single layer, nanocontact spin torque
devices. A collimated spin wave beam was observed over a range of
currents and applied field angles, with the direction of the beam
determined by the applied field azimuthal angle.  The interplay of the
applied field and the Oersted field act to form a spin wave corral,
effectively trapping excitations under the nanocontact except along
the direction where the oscillation frequency matches available
propagation states.

\acknowledgments{The authors thank Bengt Fornberg and Keith Julien for
  suggestions involving the numerical method used here.  We also thank
  the University of Colorado at Boulder's Applied Mathematics
  Department for the generous use of its computational lab.}

%=========================================================================

\bibliographystyle{apsrev}
%\bibliography{jabref}

\renewcommand{\theequation}{A.\arabic{equation}}
\setcounter{equation}{0}
\appendix*
\subsection*{Appendix}
\label{sec:methods}

The inverse Fourier transform of eq.\ \eqref{eq:1} gives
\begin{equation}
  \label{eq:16}
  \begin{split}
    \vec{m}^*_\perp &\equiv \int_0^{2\pi}\int_0^{1}
    \vec{u}_\perp(r',\phi',\tau) [K_L(R) + K_R(R)] r' \, dr'\, d\phi', \\
    &R(r',\phi';r,\phi) \equiv \sqrt{r^2 + r'^2 - 2 r r'
      \cos(\phi-\phi')} ,
  \end{split}
\end{equation}
where lengths have been normalized by the contact radius $r_*$, $R$ is
the distance between the reference $(r',\phi')$ and source $(r,\phi)$
points, $\tau$ is time, and we have assumed the thin film limit with
$\pd{\vec{u}}{z} = 0$.  The kernels $K_L$ and $K_R$ are associated
with the left and right interfaces, respectively.  Their general
expression is
\begin{equation*}
  \begin{split}
    K(r) = &\frac{a}{r} \int_0^\infty \frac{J_0(k) k \, \ud k}{\kappa
      \coth(l\kappa /r) + r/b}, ~ b=D/(w_0 r_*), \\
    \kappa \equiv &[k^2 + (rd)^2]^{1/2}, ~ d=r_*/l_{sf}, ~ l=l'/r_* , \\
    a = &r_*(\pm Q_{zz} + w_0
    m_\parallel)/(2\pi D).
      \end{split}
\end{equation*}
\begin{figure}
  \centering
  \includegraphics[width=0.5\columnwidth]{%figures/
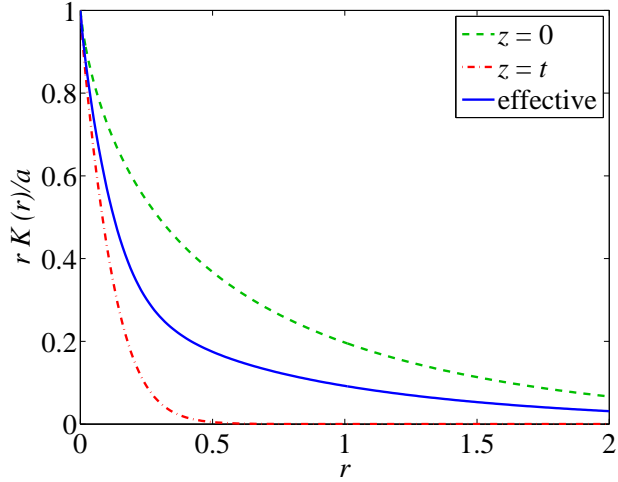}
  \caption{Lateral spin diffusion kernel $rK(r)/a$ multiplied by $r$
    to remove the singular behavior and scaled by the constant $a$ to
    compare the kernel contributions from each interface
    ($rK_L(r)/a_L$ at $z=0$, dashed curve and $rK_R(r)/a_R$ at $z=t$,
    dash-dotted curve) and their combination ($r[K_L(r)/a_L +
    K_R(r)/a_R]$, solid curve).}
\label{fig:kernel}
\end{figure}
A plot of $rK(r)/a$ is shown in Fig. \ref{fig:kernel} for the typical
parameters considered in this work.  The average magnetization
direction over the contact is
\begin{equation}
  \label{eq:13}
  \hat{u}_{\parallel} \equiv
  \frac{\vec{u}_{\parallel}}{|\vec{u}_{\parallel}|}, \quad \vec{u}_\parallel
  \equiv \frac{1}{\pi} \int_0^{2\pi} \int_0^1 \vec{u}(r',\phi',z,t) r'
  \ud  
  r' \ud \phi',  
\end{equation}
which we use as the orientation of the longitudinal spin accumulation.
The total spin accumulation is
\begin{equation}
  \label{eq:14}
  \vec{m}^* = \vec{m}_{\perp}^* + [m_z(0)+m_z(t)]
  \hat{u}_\parallel, ~  
  \vec{u}_\perp = \vec{u} - (\vec{u} \cdot \hat{u}_\parallel)
  \hat{u}_\parallel , 
\end{equation}
where the longitudinal spin accumulation is the sum of the
contributions from each interface $z=0$ and $z=t$.

For a uniform applied field of the form $\vec{h}_* =
h_*[\sin(\theta_*) \cos(\psi_*), \sin(\theta_*) \sin(\psi_*),
\cos(\theta_*)]$, the dispersion relation for exchange spin waves in a
thin film is
\begin{equation}
  \label{eq:5}
  \begin{split}
    \omega^2 = ~ &[\eta k^2 + h_* \cos(\theta_e - \theta_*) -
    \cos^2(\theta_e)] \times \\
    &[\eta k^2 + h_* \cos(\theta_e - \theta_*) -
    \cos(2\theta_e)],
  \end{split}
\end{equation}
where $\theta_e$ is the equilibrium magnetization polar angle
satisfying $h_* \sin(\theta_e - \theta_*) - \frac{1}{2}
\sin(2\theta_e) = 0$.

We briefly discuss the numerical method we have used to solve eq.\
\eqref{eq:6}.  
% A detailed presentation and error analysis will be given elsewhere.
The polar coordinate system is a particularly efficient and accurate
choice for nanocontact simulations.  The discretization we use is
non-uniform in radius (``inner'' and ``outer'' grids) and uniform in
angle
\begin{equation*}
  \begin{split}
    r_i &= q(i-1/2) \equiv \frac{1}{2}(dr_{in}+dr_{out}) (i-1/2) +
    \\
    & \quad \frac{1}{2} (-dr_{in}+dr_{out}) w
    \ln[\cosh(\frac{\hat{i}-i+1/2}{w})/
    \cosh(\frac{\hat{i}}{w})] , \\
    \phi_j &= -\pi\! +\! (j\!-\! 1)d\phi, \, d\phi = \frac{2\pi}{M},
    \, i\! =\! 1,\ldots,N, ~j\! =\! 1,\ldots, M,
  \end{split}
\end{equation*}
where $\hat{i}$ and $w$ are parameters determining the location and
width of the smooth change from the fine inner grid spacing $dr_{in}$
to the coarser outer grid spacing $dr_{out}$.  The advantage of this
discretization is that we can solve on a uniform computational grid
$(i-1/2,\phi_j)$, but the physical grid $(r_i,\phi_j)$ is clustered in
and around the point contact where the interesting dynamics occur.
The outer grid supports the propagation of spin waves of the
appropriate wavelength away from the point contact.  We choose a
domain large enough $0 < r < L$, $L \gg 1$ so that wave reflections
off the boundary do not affect the solution inside the point contact.
For comparison, our simulations on a circular domain with diameter
$2Lr_* = 2\cdot 100\cdot 40$ nm $=8 ~ \mu$m involve $MN = 32\cdot 502
= 16,064$ grid points.  A nonuniform grid in the Cartesian coordinate
system covering the same domain would involve $4N^2 = 1$ million
gridpoints while a uniform grid would require 16 million gridpoints.

The boundary condition for eq.\ \eqref{eq:6} is the Neumann condition
$\pd{\vec{u}}{r}(r=L,\phi,\tau) = 0$.  We approximate radial
derivatives using sixth order finite differences and the angular
derivative $\vec{u}_{\phi \phi}$ is calculated using a pseudospectral,
fast fourier transform method.  An explicit Runge-Kutta
$4^{\text{th}}$ order time stepping method is used to advance equation
\eqref{eq:6} forward in time.  To avoid severe time step restrictions
due to the small grid spacing near the origin, we apply a smooth,
radial grid dependent angular mask $g_i(k)$ at every time step that
filters out numerically induced small wavelengths near the origin
\cite{Fornberg1995}.  The mask applied to positive angular mode $k$ at
the radial grid point $r_i$ takes the form
\begin{equation*}
  g_i(k) = \frac{1}{2} + \frac{1}{2}\tanh\left(\frac{k_i - k}{\Delta
      k}  \right), ~ k_i = \frac{k_0 r_i}{2 r_1},
\end{equation*}
and is evenly extended to negative $k$ values.

Numerical parameters we use are: $dr_{in}= 0.05$, $dr_{out} = 0.25$,
$w = 10$, ($\hat{i} = 126$, $M = 32$, $N = 342$) for Figs.\
\ref{fig:fmr_angle} and \ref{fig:fmr_current}, ($\hat{i} = 273$, $M =
64$, $N = 455$) for Figs.\ \ref{fig:modes_angle} and
\ref{fig:modes_I}, $L = 60$, $k_0 = 3$ (wavenumber cutoff in angular
mask at $r=r_1$), and $\Delta k = 1$ (width of wavenumber cutoff in
angular mask).  We find no significant change in the results for more
accurate grids and filtering parameters.

The convolution $m_\perp^*\{\vec{u}_\perp\}$ in eq.\ \eqref{eq:16} is
evaluated on the computational grid using Simpson's rule in both the
angular and radial directions.  To deal with the removable singularity
when $R=0$ in eq.\ \eqref{eq:16}, we subtract off the small $R$
behavior
\begin{equation*}
  K(R) = af(R)/R \sim a(1/R + \ln(R)/b + G) + o(1), \quad R \ll 1,
\end{equation*}
where $G$ is a constant.  The resulting integral, at one interface,
that we evaluate numerically is
\begin{equation*}
  \begin{split}
    \vec{m}_\perp = &\int_0^{2\pi}\!\! \int_0^1 \frac{a
      f(R) \vec{u}_\perp' - a[1+R \ln(R)/b] \vec{u}_\perp}{R} r' \, \ud
    r' \, \ud \phi' + \\
    & a 4 E(r^2)\vec{u}_\perp + \frac{a}{b} \vec{u}_\perp \int_0^{2\pi} \int_0^1
    \ln(R) r'
    \, \ud r', \\
    \vec{u}_\perp' \equiv &\vec{u}_\perp(r',\phi',\tau), ~ \vec{u}_\perp \equiv
    \vec{u}_\perp(r,\phi,\tau) ,
  \end{split}
\end{equation*}
where $4 E(r^2) = \int_0^{2\pi} \int_0^1 r'/R \, \ud r' \, \ud \phi'$
and $E$ is the complete elliptic integral of the second kind.

\end{document}